\documentclass[usegraphicx]{mn2e}

\usepackage{graphicx}
\usepackage{subfigure}
\usepackage{longtable}\usepackage{lscape}
\usepackage{times}
\usepackage{epsfig}
\newcommand{\integral}{{\it INTEGRAL}}
\newcommand{\rxte}{{\it RXTE}}

\newcommand{\beppo}{{\it BeppoSAX}}
\newcommand{\cgro}{{\it CGRO}}
\newcommand{\asca}{{\it ASCA}}
\newcommand{\cyg}{{Cyg~X-1}}
\def\ergcms{erg cm$^{-2}$ s$^{-1}$ }

\title[Magnetic field in \cyg]{The magnetic field in the X-ray corona of Cygnus X-1
\thanks{Based on observations with {\it INTEGRAL}, an ESA  project with instruments and science data centre funded by ESA member states 
(especially the PI countries: Denmark, France, Germany, Italy,  Switzerland, Spain), Czech Republic and Poland, 
and with participation  of Russia and the USA.}}
\author[M. Del Santo et al.]{M. Del Santo$^{1}$, J. Malzac$^{2,3}$, R. Belmont$^{2,3}$, L. Bouchet$^{2,3}$, G. De Cesare$^{1}$\\
$^{1}$Istituto Nazionale di Astrofisica, IAPS, via Fosso del Cavaliere 100, 00133 Roma, Italy \\
$^{2}$Universit\'e de Toulouse, UPS-OMP, IRAP, Toulouse, France\\
$^{3}$CNRS, IRAP, 9 Av. Colonel Roche, BP 44346, 31028 Toulouse Cedex 4, France}

\begin{document}
\date{Accepted 2012 December 06. Received 2012 December 05; in original form 2012 July 05}

\pagerange{\pageref{firstpage}--\pageref{lastpage}} \pubyear{2012}

\maketitle

\label{firstpage}

\begin{abstract}
The different electron distributions in the hard and soft spectral states of black-hole binaries  could be caused by kinetic processes 
and changing because of varying physical conditions in the corona.
In presence of a magnetic field in the corona, the electron distribution can appear thermal, even when acceleration mechanisms would produce non thermal distributions.
This is due to fast and efficient thermalization through synchrotron self-absorption.
 In this paper, we have analyzed data from  6 years of observations of Cygnus X-1 with the \integral\ observatory
 and  produced 12 high-quality,  stacked broad-band hard X-ray spectra representative of the whole range of spectral shapes observed in this source.  
 We then fit these spectra with hybrid thermal/non-thermal Comptonization models and study the evolution of the physical parameters of the accretion flow across the spectral transition. 
 In  particular, we use the {\sc belm} model to constrain the magnetic field in the corona through its effects on the coronal emission. 
Indeed, the hot electrons of the X-ray corona produce soft (optical-UV) synchrotron radiation which is then Comptonized and may affect the temperature of the electrons (and thus the slope of the X-ray spectrum) through Compton cooling.  We find that in the softer states, the emission is dominated by Comptonization of the disc photons and the magnetic field is at most of the order of $10^{6}$ G. 
In the harder states, the data are consistent with a pure synchrotron self-Compton model, although a significant contribution of Comptonization of disc photons may not be excluded. 
If the non-thermal excess observed above a few hundred keV in the hard state is produced in the same region as the bulk of the thermal Comptonization,
we obtain an upper limit on the coronal magnetic field of about $10^{5}$ G.  If, on the other hand, the non-thermal excess is produced in a different location 
(such as the jet or a different part of the corona), the constraints on the magnetic field in the hard state are somewhat relaxed and the upper limit  rises to  $\sim 10^{7}$  G. 
We discuss these constraints in the context of current accretion flow models.
\end{abstract}


\section{Introduction}\label{sec:intro}

Black hole binaries (BHBs), and \cyg\ in particular, show dramatic spectral variability. 
The most dramatic variations observed in BHBs are spectral transitions which occur, for most of these sources,
on the time scales from weeks to days \cite{zdz02}.
In the simplest classification, BHBs, are observed in two main spectral states:
in the hard state (HS) the source emits essentially  radiation peaking around 100 keV \cite{gier97,ZG04}, 
while in the soft state (SS) the X-ray spectrum is a prominent  $\sim$1 keV black-body plus a power-law tail (Zdziarski 2000 and ref. therein).
The HS spectrum is believed to originate from thermal Comptonization in a hot electrons dominated cloud (the so-called "corona"; Eardley, Lightman \& Shapiro 1975;  Sunyaev \& Titarchuk 1980).

The black-body dominated spectrum observed in SS was associated with the optically
thick accretion disc  \cite{ss73}, while the origin of the weak steep power-law component is still debated  \cite{zdz00}.
In addition spectra with properties that are intermediate between that of the SS and HS are observed when the sources switches from one of the two stable state to the other. 
Terms such as "hard intermediate", or "soft intermediate" are often used to describe different flavors of those intermediate state \cite{belloni05} but there is no unique classification (see McClintock \& Remillard 2006).

 The spectra of BHBs consist in several components.
 In the HS of \cyg, the electron temperature  and the Thomson optical depth of the Comptonizing plasma were found to be typically 
 $kT_{\rm e}$ = 50--100 keV and  $\tau$ = 1--3  \cite{gier97,frontera01}. 
 Although the scattering electrons have a predominantly Maxwellian energy distribution in the HS, there are indications 
 that the electron distribution may have some high-energy tail  i.e., it is hybrid, thermal/non-thermal. 
The first evidence has been provided by \cgro/COMPTEL observations at  E $>$ 500 keV of \cyg\ in hard state  \cite{mac94,mac02}.
A slight indication was also inferred by \cgro/OSSE observations of GX 339--4 \cite{ward02}.
Then, with the \integral\ mission, this feature has been observed many times during hard and/or intermediate/hard states in a number of BHBs and BH candidates (BHC; see Del Santo 2012
for a review on this topic), 
such as 1E 1740.7--2942  \cite{bouchet09}, GX 339--4 \cite{delsanto08,joinet07,droulans10}, 
Cyg X--1 (Malzac et al. 2006; Cadolle Bel et al. 2006; Jourdain, Roques \& Malzac 2012; Zdziarski, Lubi\'nski \& Sikora 2012).

Alternatively, it is possible that the non-thermal excess and thermal Comptonization component are not produced by the same electron distribution but in separate regions of the corona with different physical properties (Malzac 2012). Also, radio observations of Cygnus X-1 have resolved a compact relativistic jet (Stirling et al. 2001) and it was  suggested that the high energy excess may be a component arising from it  (Rahoui et al. 2011). This idea has been also proposed following the hard X-ray polarization measurements by Laurent et al. (2011) obtained with the IBIS telescope in Compton mode.  These authors found that, while the 250--400 keV spectrum is consistent with emission dominated by Compton scattering by thermal electrons and are weakly polarized (Pf  $<$ 20\%), the second spectral component seen in the 400 keV--2 MeV band is strongly
polarized (Pf = 67$\pm$ 30\%). They have argued that the MeV excess is likely to be produced in jet through synchrotron emission in a very coherent magnetic field.  Such a high degree of polarization is indeed difficult to achieve through inverse Compton emission in the corona. 
Recently Jourdain et al. (2012) reported on a similar result by using the SPI telescope. They found that  above 230 keV the \cyg\ emission is indeed polarized,
with a mean polarization fraction of  76$\pm$ 15\%.

The shape of the MeV excess however, appears difficult to reproduce in the jet scenario (Zdziarski, Lubi\'nski \& Sikora 2012). Such a scenario would also imply that the MeV excess in the HS  and the non-thermal power-law emission observed  in the SS have a completely different nature since the radio jet is quenched in the SS.  We note that although it is very  plausible that the non-thermal excess originates in the jet, it appears very unlikely  that the jet emission  could dominate the entire high energy spectrum of \cyg\ (see e.g. Malzac, Belmont \& Fabian 2009 and references therein).

The X/$\gamma$-ray  SS spectrum of \cyg\  has been studied extensively
by simultaneous observations with \asca, \rxte, \beppo, and \cgro\ during summer 1996 \cite{disalvo01,frontera01}.
In addition to the dominating black-body, a long power-law like tail extending up
to 10 MeV was discovered  \cite{mac02}. 
The high energy spectrum is well described by single Compton
scattering off electrons having a nearly power-law distribution (Gierli\'nski et al. 1999, hereafter G99; Frontera et al. 2001). 

The different spectral states are usually understood in terms of changes in the geometry of the accretion flow. 
According to a popular scenario (see e.g. Done, Gierli\'nski \& Kubota 2007 for an exhaustive review), in the SS a geometrically thin accretion disc (the standard Shakura-Sunyaev disc)
extends down to the last stable orbit and is responsible for the dominant black-body emission. This disc is the source of soft seed photons for
Comptonization in small active coronal regions located above and below the disc. The magnetic field lines rise above the accretion
disc through magnetic buoyancy, transporting a significant fraction of the accretion power into the corona where it is then dissipated
through magnetic reconnection  \cite{galeev79}.
This leads to particle acceleration in the corona. A population of high-energy electrons is formed which then cools down by up-scattering
the soft photons emitted by the disc. This produces the high-energy non-thermal emission (see e.g. G99; Zdziarski et al. 2002).
In the HS, the standard geometrically thin disc  is truncated at distances ranging from a few
tens to a few thousand gravitational radii from the black hole. In its inner parts, the accretion flow takes
the form of a hot geometrically thick, optically thin disc (Esin, McClintock \& Narayan 1997). In this hot accretion flow, the  electrons are predominantly heated by Coulomb interaction with a population of hot ions and cool down by  Comptonizing their own synchrotron emission and/or soft photons from the accretion disc.

The different electron distributions in HS and SS  could be caused by kinetic processes 
and changing because of varying physical conditions in the corona (Malzac \& Belmont 2009, hereafter MB09; Poutanen \& Vurm 2009, hereafter PV09).
These authors showed that in presence of a magnetic field in the corona,
 the electron distribution can appear thermal,
 even when acceleration mechanisms would produce non thermal distributions.
 This is due to fast and efficient thermalization through synchrotron self-absorption as first pointed out by Ghisellini, Guilbert \& Svensson (1988).
MB09 and PV09, presented a rough 'fit by eye' of the average CGRO data of \cyg\ with this synchrotron boiler model which already provided estimates of the magnetic field in the hard state under the assumption that the non-thermal excess is produced by electrons in the same zone as the Maxwellian component. In this paper, we will use the same model to perform a statistical fit of 
\integral\  data for the whole range of observed spectral shapes and also considering the possibility that the non-thermal excess in the hard state may have a different origin. 
 
As one of the brighter galactic hard X-ray source, \cyg\ is a prime target for the \integral\ mission \cite{wink03}.
It was extensively observed during the Performance Verification (PV) Phase of the mission.
The on-board instruments offer an unprecedented simultaneous broad-band spectral coverage,  ranging from 3 keV to several MeV. 
Thus a large amount of observing time (open time and core programme for a total of about 7 Ms) have been dedicated to this target by \integral.
A number of studies of the intermediate spectral states have been performed (Malzac et al. 2006; Cadolle Bel et al. 2006);
an X-ray flare during the hard state appears to be coincident with the TeV emission  detected by MAGIC (Albert et al. 2007; Malzac et al. 2008).

Recent papers have also provided a refined estimation of the distance (d=1.86 kpc; Reid et al. 2011) and the black-hole mass of \cyg\ (14.8 M$_{\odot}$; Orosz et al. 2011).

In this  work, we present the long term behaviour and a spectral variability study of \cyg\ using the whole \integral\ data base available until Spring 2009. Rather than following the evolution of the source chronologically, the aim was to extract high quality data representative of the source in a given 'spectral state'. For this purpose, we stacked all the pointings with similar hardness ratios to produce 12 averaged spectra spanning the whole range of observed spectral shapes. The details of how this was done are described in Section~\ref{sec:data}. 
These 12 broad band (3 keV--1 MeV) spectra were then fit with two different hybrid thermal Comptonization models namely  {\sc eqpair} \cite{coppi99} and {\sc belm}  (Belmont, Malzac \& Marcowith 2008; hereafter B08) that are described in Section~\ref{sec:models}. 
This allows us to constrain  the physical conditions in the corona and determine how the physical parameters change during the spectral evolution. In particular, fits with {\sc belm} sets quantitative constraints on the strength of the magnetic field in the corona of Cygnus X-1 for the first time. These results are described and discussed in Section~\ref{sec:results}.

\begin{figure}
\centering
\includegraphics[height=8cm,angle=+90]{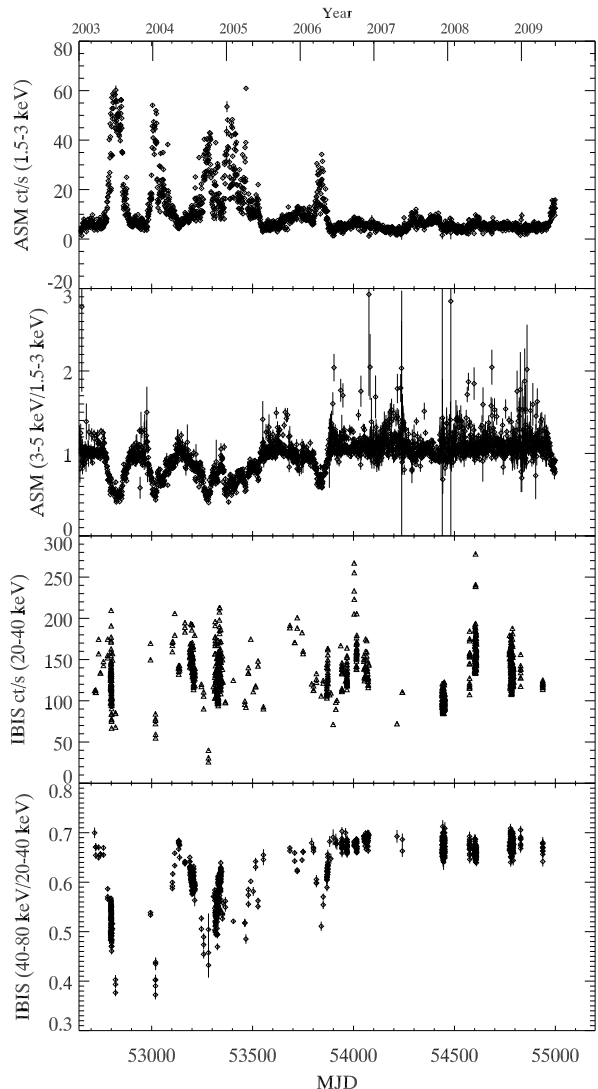}
\caption{From top to bottom: ASM count rate in 1.5-3 keV, ASM hardness ratio (3-6 keV/1.5-3 keV),  IBIS/ISGRI count rate in 20-40 keV,  IBIS/ISGRI hardness (40-80 keV/20-40 keV).
 \label{fig:lc}}
\end{figure}

\section{Observations and data analysis}\label{sec:data}

\begin{figure}
\includegraphics[height=6cm,width=9cm]{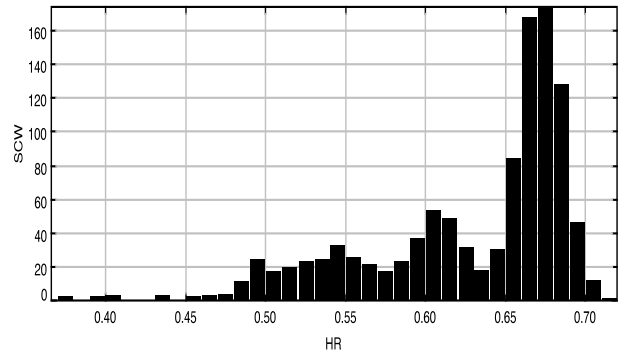}
\vspace{-1cm}
\caption{Number of the \integral\ pointings considered in this paper vs IBIS/ISGRI hardness ratio (40-80 keV/20-40 keV). \label{fig:histo}}
\end{figure}

We have analyzed six years of  \cyg\  observations performed by \integral\ until April 2009 (revolution 794) with the three coded-mask telescopes IBIS \cite{ubertini03}, 
SPI \cite{vedrenne03} and JEM-X \cite{lund03}.
We started from March 2003 (revolution 52), thus we excluded the performance verification phase observations (1 Ms) due to some peculiar instrument settings.
Results on \cyg\ obtained during the calibration phase have been extensively reported (Bouchet et al. 2003; Pottschmidt et al. 2003; Bazzano et al. 2003). 

For the purpose of spectral analysis,  we have built our data-set by selecting all IBIS observations including 
\cyg\ in a partially coded FOV of $15^{\circ}  \times 15^{\circ}$ where the instrument response is well known.
Based on this selection, the total number of IBIS and SPI science windows (SCWs) results as 1024
(Tab. \ref{tab:log}) for a total exposure time of about 3.3 Ms (a single SCW can last from 800 up to 4000 seconds).
The total SPI exposure time is essentially the same as that of IBIS, even though a further analysis excluding SPI problematic
pointings (i.e. solar flares, radiation belts) leads to a slightly lower SPI effective exposure time than IBIS.
Because of the smaller FOV of the X-ray telescope JEM-X, the related selected pointings  (FOV of $3.5^{\circ}  \times 3.5^{\circ}$ has been considered) 
are 364 (see Tab. \ref{tab:log}). In particular,  44 SCWs for JEM-X2, operating at the beginning of the mission and 320 for JEM-X1 operating since revolution 181.

The IBIS scientific analysis, focused on the low energy detector ISGRI \cite{lebrun03}, and the JEM-X analysis have 
been performed using the \integral\  off-line analysis software, OSA v.9 \cite{gol03,cour03}.  
The IBIS/ISGRI light curves (Fig. \ref{fig:lc}) have been extracted in two energy bands (20$-$40 keV, 40$-$80 keV). 
JEM-X and IBIS/ISGRI spectra, extracted SCW by SCW, extend from 3 to 20 keV and from 20 to 500 keV, respectively.

To account for change over time of the energy response, IBIS is calibrated in different periods using dedicated observations of the
Crab. Then for each Crab observation an ancillary response file (ARF) is
estimated and associated with a given interval of revolutions in the \integral\ data  (see IBIS User Manual). 
Since spectra in our data analysis (see Section \ref{sec:vary}) often correspond to observations collected in different periods, i.e.
associated with different ARF files, we evaluated for each averaged spectrum an ARF file obtained by a weighted mean of the effective areas in
the different periods involved.

\begin{table}
\begin{center}
\caption{Observations log of the twelve final data-set used for spectral modeling. The IBIS harness ratio intervals (HR), number of pointings (SCW) with IBIS and SPI and with JEM-X (within brackets), 
IBIS and SPI observing exposures  (Exp1), JEM-X exposure (Exp2) and the power-law slope ($\Gamma$) of the  30--100 keV IBIS/ISGRI spectra are shown.}\label{tab:log}
\begin{tabular}{c|cccc} 
\hline
HR & SCW (JEM-X) & Exp1$^{\dag}$ &Exp2$^{\ddag}$ &  $\Gamma$ \\
       &  number          &  [ks]    &   [ks]     &          \\
\hline

0.37-0.41 & 7 (1)  & 15.4 & 2.2 & 2.6$^{+0.04}_{-0.03}$\\
0.43-0.48 & 12 (2)  & 36.8 & 6.5 & $2.32\pm 0.02$ \\
0.48-0.50 &  35 (9)   &      111.5  & 27.8 &      $2.29\pm 0.01$               \\
0.50-0.53 & 59 (8)         &  181.1     &     22.9       &  $2.23\pm 0.02$               \\
0.53-0.56 & 82 (11)         &  236.8     &     34.0       &  $2.17\pm 0.02$               \\
0.56-0.58 & 38 (19)         &  101.0    &     51.1      &  $2.08\pm 0.02$               \\
0.58-0.61 & 113 (40)      &  343.4    &  113.7 &   $2.02\pm 0.02$               \\
0.61-0.64 & 97 (27)      &  287.7    &  80.3 &   $1.98\pm 0.02$               \\
0.64-0.66 & 114 (29)      &  944.2    & 34.7 &   $1.81\pm 0.02$               \\
0.66-0.68 & 340 (72)      &  781.4    & 147.8 &   $1.79\pm 0.02$               \\
0.68-0.70 & 157 (16)      &  402.0   &  34.2 &   $1.74\pm 0.02$               \\
0.70-0.72 & 9 (4)      & 26.1    &  8.0 &   $1.73\pm 0.02$               \\

\hline
\end{tabular}
\end{center}
\end{table}

\subsection{SPI data treatment}
 The signal recorded by SPI camera is composed of contributions from sources in the FOV   convolved with the instrument response function plus the background. 
A system of equations is to be solved to determine sources and background intensities. 
 In order to reduce the number of uncertainties and hence to maximize the signal-to-noise ratio of each sources, a few assumptions are introduced. 

The background  count rates of the 19 Ge detectors (uniformity maps) are assumed to be stable on timescale of $\sim$6 months,
 while the global normalization factor is determined on $\sim$6 hours timescale. 
 For the sources, the timescales are chosen in function of the source intensity and temporal behavior:
 the faint sources are considered as constant. More information and details on SPI data reduction can be found in Bouchet et al. (2008) and Bouchet, Roques \& Jourdain (2011).

At energy around 1 MeV, high energy particles saturate the electronics and can generate false events. 
The trigger signal issued by the Pulse Shape Discriminators is used to select events between 650 keV and 2.2 MeV as explained in Jourdain \& Roques (2009).

In order to maximize the signal-to-noise ratio, we used a catalog containing sources detected above 2$\sigma$ in the whole time interval spawned by the data. 
The catalogue contains 10, 6, 5 and 1 sources in the 25-50 keV, 50-100 keV, 100-300 keV and 300-600 keV energy bands, respectively. 
In addition \cyg\ is time-variable up to $\sim$400 keV, while above $\sim$400 keV, the signal-to-noise ratio is better  when assuming \cyg\ as constant.
Above 500 keV, we can derive a better uniformity map using high latitude exposures (telescope pointing axis satisfying $|b|$ $>$ 30$^\circ$) 
as they contain no significant source emission \cite{bouchet10}. Comparison of the fluxes obtained with this uniformity map and one that 
is determined by the fitting algorithm (Bouchet et al. 2011) shows that the difference in determined fluxes is well smaller than the error bars and hence fluxes are fully compatible.
In this paper, we use SPI spectra from 22 keV up to 1 MeV (when possible because of the statistics).

\subsection{Long term behaviour and spectral variability}\label{sec:vary}

\begin{figure}
\centering
\includegraphics[height=8.5cm,angle=+90]{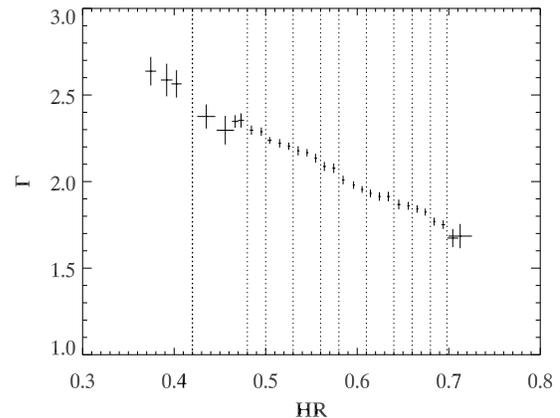}
\caption{Correlation between the spectral photon index of  the IBIS spectra fit with a simple power law in the range 30--100 keV
and the hardness ratio 40--80/20--40  keV. Dashed lines indicate the twelve data groups used for the final spectral analysis (see text).
\label{fig:gamma}}
\end{figure}

As a first step in characterizing the spectral variability (i.e. spectral states) of \cyg, we computed the hardness ratio (HR) defined as the count rate ratio 40--80 keV to 20--40 keV
 for each one of the 1024 IBIS pointings (Fig. \ref{fig:lc}, panel 4).
\cyg\ was  most of the time in HS so that observations in hard states  are significantly more numerous than the ones in soft states (Fig.  \ref{fig:histo}).
During the first three years of the \integral\ monitoring, \cyg\ showed a very variable activity in term of both flux variation and spectral transition.
Since middle of 2006, the source entered  in an almost steady hard state, characterized by flux variability not combined with simultaneous spectral variability (see the two
flares around MJD$\sim$54000 and MJD$\sim$54600 and the corresponding flat HR in Fig. \ref{fig:lc} panel 3 and 4, respectively).
For comparison with the soft X-rays, we plot the \rxte/ASM long term behaviour, as well as the hardness ratio (Fig. \ref{fig:lc}, panels 1 and 2).

We have fitted individually all 1024 IBIS/ISGRI spectra from  20 keV to 80 keV with a simple power law model and verified that at a SCW level 
IBIS spectra in this band are described by a power-law with a null hypothesis probability higher than 5\%. 
Therefore, the hardness ratio we have defined can be used as a spectral state indicator.

The 1024 IBIS/ISGRI by-pointing spectra  have been first  averaged within each HR bin (HR bin-size=0.01), resulting in 31 spectra covering the range HR=0.37$\div$0.72. 
These spectra have been fitted in the range 30--100 keV with a simple power-law.
The correlation photon index vs. HR is shown in Fig.~\ref{fig:gamma}.
Then,  for the second step in averaging our IBIS spectra,  in order to obtain a better statistics at high energy (above 100 keV), we have used two combined criteria.
We averaged data from neighboring HR bins with compatible (90\% confidence level) 30--100 keV power-law slopes (first criterium; see Fig. ~\ref{fig:gamma}) 
until the error on $\Gamma$ is reduced down to at least 0.02 (second criterium; Tab. \ref{tab:log}). 
At the end we have obtained 12 IBIS spectra collected with distinct shape spanning most of the range of observed spectral diversity of \cyg.

The JEM-X spectra have been then averaged following the IBIS selection.
We checked the JEM-X spectral variability inside our IBIS HR bins and found that spectra inside
the HR bin we have chosen are compatible with a constant slope in the JEM-X band.
The SPI  spectra have been extracted from the same 12 IBIS SCW groups.

\section{Models}\label{sec:models}

The twelve averaged broad-band spectra have been fitted with XSPEC v. 11.3.2, using two Comptonization models, namely {\sc eqpair} (Coppi 1999, G99) and {\sc belm} (B08, MB09), plus disc reflection and relativistic iron line emission. The best-fit parameters of the twelve averaged broad-band spectra are shown in Tab.~\ref{tab:fit} ({\sc eqpair}) and Tab.~\ref{tab:belm} ({\sc belm}). 
All quoted errors are at 90\% confidence level.

\subsection{Coronal emission models}
{\sc eqpair} and {\sc belm} model the emission from a homogeneous, isotropic, spherical cloud of fully ionized matter in interaction with the ambient radiation field. Not only do they compute the observed photon spectrum of  the corona, but also they give the energy distributions of leptons in the corona. Those results are computed self-consistently according to the following microphysical processes: lepton-lepton Coulomb collisions, Compton scattering, photon-photon pair production, pair annihilation, self-absorbed bremsstrahlung radiation, and self-absorbed synchrotron radiation (the latter for {\sc belm} only). 
The leptons distributions are generally found to be composed of a low-energy thermal population and an additional high-energy non-thermal population, so that those kind of models are usually called {\it hybrid} models. The main physical parameters that characterize such corona are its typical size, density, magnetic field, and how it is energized. However, the spectral properties depend more explicitly on non-dimensional combinations of these quantities. 

For instance, the dependence on the absolute coronal size is negligible for most parameter regimes, and the radius is set to $R=5\times10^{7}$ cm, 
which corresponds to a region of about $R\approx 20 R_g$ for a $15 M_\odot$ black hole. 

The proton density $n_{\rm p}$ is represented by the proton Thomson optical depth $\tau_{\rm p} = n_{\rm p} \sigma_{\rm T} R$, where $\sigma_{\rm T}$ is the Thomson cross section. 
The Thomson optical depth $\tau_{\rm p}$ provided by the two models is related to the ionization electrons only. 
The total optical depth ($\tau_{\rm tot}$) is the sum of the optical depth of $e^{+}e^{-}$ pairs plus $\tau_{\rm p}$.  

The source luminosity $L$ is described by the non-dimensional compactness parameter \cite{LZ87}: 

\begin{equation}
\label{eq:comp}
l = \frac{ \sigma_{T} } { m_{e} c^{3}} \frac{L}{R}
\end{equation}

where $m_e$ is the electron mass, and $c$ the speed of light.

Photon escape can only be dealt with exactly by Monte-Carlo or complete radiation transfer simulation. Instead, it is modeled here through an approximate escape 
probability formalism (e.g. Lightman, Zdziarski \& Rees 1987; Lightman \& Zdziarski 1987; Stern et al. 1995). {\sc eqpair} uses an escape probability based on pure photon scattering, whereas the escape 
probability used in the present version of {\sc belm} is based on both photon scattering and absorption (Sobolev 1974; VP09), 
which can lead to significant differences for low energy photons and large optical depths. Although results with the two models can give slightly different 
results for a given set of parameters, the spectral dependence on parameters is identical in both models.

To compensate for losses associated to photon escape, energy can be injected in the corona (i) directly as external photons from the accretion disc, or indirectly to the leptons populations through two channels: (ii) thermal heating or (iii) non-thermal acceleration. 
\begin{enumerate}
\item External photons are assumed to originate from a optically thick, multi-temperature accretion disk extending from the minimal stable orbit of a non-rotating 
black hole $R_{\rm in}=6 R_g$ out to $R_{\rm out} = 10^3 R_g$. The emitted multi-color black-body spectrum is then essentially determined by the temperature $T_{\rm in}$ at the inner radius.
As for the source luminosity, the power injected as soft photons is characterized by the {\it soft} compactness: $l_{\rm s}$ (also named $l_{\rm bb}$ in {\sc eqpair}) . 
\item Thermal heating provides energy to the pool of thermal leptons with a compactness parameter $l_{\rm th}$. 
\item Non thermal processes such as magnetic reconnection and/or shock acceleration are expected to generate power-law like distributions of particles. This is modeled as an injection of leptons 
with a power-law distribution $f(\gamma)\propto \gamma^{-\Gamma_{\rm inj}} $ characterized by the minimal and maximal Lorentz factor $\gamma_{min}=1.3$ and $\gamma_{max}=10^3$ between which the acceleration takes place, the power-law index $\Gamma_{\rm inj}$, and the total injected power described by the compactness $l_{\rm nth}$. 
\end{enumerate}
$l_{\rm h}=l_{\rm th}+l_{\rm nth}$ corresponds to the total power supplied directly to the plasma.
In steady state, all injected power goes out as photons and $l=l_{\rm s}+l_{\rm th}+l_{\rm nth}$. As far as the compactness $l$ does not reach too large values ($l < 100$), 
the spectral shape is rather insensitive to the total compactness and to the absolute values of $l_{\rm s}$, $l_{\rm th}$, and $l_{\rm nth}$. However, it depends on the channel used to injected this power, i.e. on ratios such as $l_{\rm h}/l_{\rm s}$ and $l_{\rm nth}/l_{\rm h}$.
For this reason, it is customary to fix $l_{\rm s}$ to some reasonable value when fitting data with  {\sc eqpair}.
 
 The main difference between these two models is that in addition to all physical process mentioned earlier, 
the {\sc belm} model allows for self-absorbed synchrotron radiation resulting from the gyro-motion of leptons around a tangled magnetic field, $B$  (B08). 
The magnetic field intensity is set through the magnetic compactness parameter: 
\begin{equation}
\label{eq:magnetic}
l_B=\frac{\sigma_{\rm T}}{m_e c^2} R \frac{B^2}{8\pi}\quad
\end{equation}

Since the shape of the spectrum depends on the ratio  $l_B/l$ rather than on the absolute value of $l_B$ we computed the {\sc belm} code
leaving  $l_B/l_{\rm h}$ as free spectral parameter.

Also, in the {\sc belm} code the thermalization process is treated self-consistently 
while in {\sc eqpair} the electron distribution is assumed to be purely thermal at low energy (see MB09 for details).

The computation time  with {\sc belm} is comparatively longer than with {\sc eqpair} and the former cannot be used for inline fitting. 
Instead, a table of spectra was computed and used in XSPEC as a table model.

\subsection{Reflection on the accretion disc}

Compton reflection has been modeled by using the viewing-angle-dependent Green's
functions approximation obtained by Magdziarz \& Zdziarski (1995) for  an isotropic point
source (or, equivalently an optically thin corona) above a slab. We
treat $\Omega/2\pi$ as a free parameter, where $\Omega$ is an effective (i.e.,
corresponding to the observed strength of reflection) solid angle subtended by
the reflector. 
The reflection is accompanied by an Fe fluorescence K$\alpha$
line centered at an energy, $E_{\rm line}$.

When the reflection comes from a fast rotating
disc in a strong gravitational potential, Doppler and gravitational shifts
become important. These effects are usually approximated  by convolving {\it both\/} the
reflected component and the Fe K$\alpha$ line with the Schwarzschild disc line
profile of Fabian et al.\ (1989; {\sc diskline} in XSPEC). 

In calculating the relativistic distortion, we consider a
range of radii from $R_{\rm in}$=6 to $R_{\rm out}$=1000 gravitational radii .
The disc reflection component is calculated for neutral material with standard abundances. 
We assumed an inclination angle of 50$^{\circ}$.

Due to the energy resolution and line sensitivity, the iron line study cannot be performed with JEM-X. The line energy was therefore  imposed at 6.7 keV.
The inclination angle, inner radius of the disc and the disc emissivity law were fixed at the same values as in the Compton reflection model. The normalization is therefore the only free parameter for the iron line emission.

\begin{figure}
\centering
\includegraphics[height=8.8cm, angle=90]{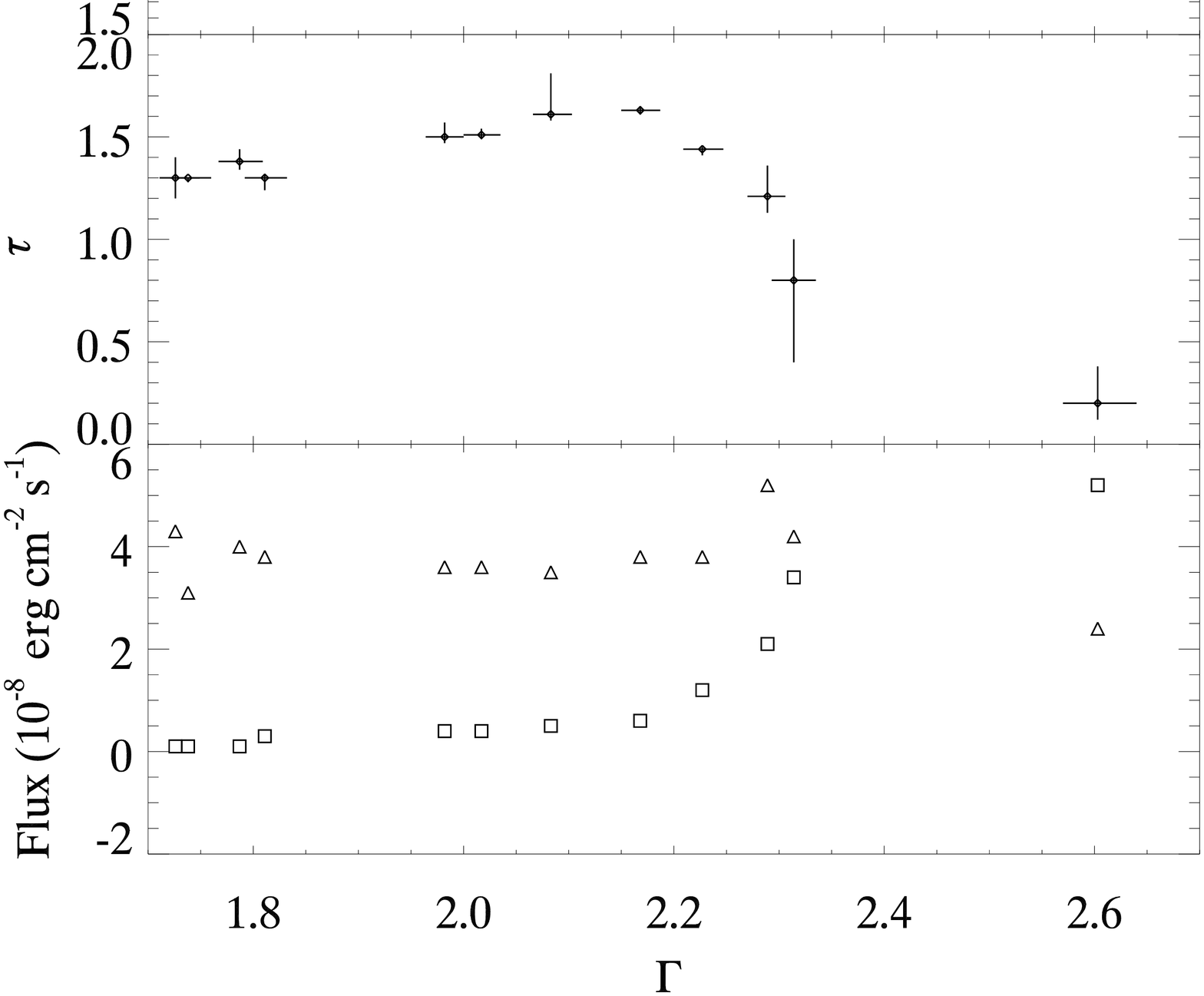}
\vspace{+0.2cm}
\caption{{\sc eqpair} spectral parameters reported in Tab. \ref{tab:fit} vs $\Gamma$ reported in Tab. \ref{tab:log}.
In the bottom panel fluxes of the  two principal components, disc black-body (squares) and Comptonization (triangles), are plotted.}
\label{fig:corr}
\end{figure}

\begin{figure}
\centering

\includegraphics[height=6.3cm]{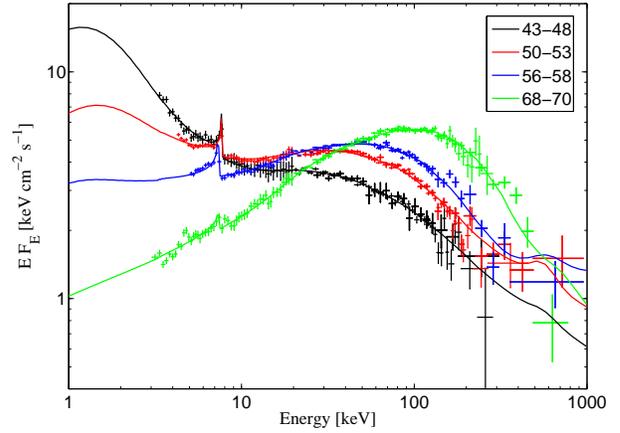}
\caption{Joint JEM-X, IBIS and SPI  energy spectra of \cyg\ during four different spectral states
fitted with the thermal/non-thermal hybrid Comptonization
model {\sc eqpair} plus {\sc diskline} (see Tab. \ref{tab:fit}).
}
\label{fig:spectra}
\end{figure}

\section{Results}\label{sec:results}

\subsection{Unmagnetized model}

As a first study, we investigate here a model with no magnetic field ($l_B$=0), so that we performed the analysis with {\sc EQPAIR}.
According with previous studies \cite{zdz02,wilms06}, we also have found that the bolometric luminosity of \cyg\ does not change significantly from SS to HS. 
Therefore, unless the size of the corona changes dramatically during the transition, the total compactness is not expected to vary much \cite{pou98}. 
However the distribution of the radiation changes dramatically. In the hard state most of the luminosity comes from the corona while in SS most of the radiation is in the black-body disc emission.
 We therefore  fixed $l_{\rm s}=10$ when fitting soft and intermediate/soft states 
 (States 1--4 in Tab. \ref{tab:fit}; G99) and $l_{\rm s} = 1$ when fitting the intermediate/hard and hard states (States 5--12;  Ibragimov et al. 2005).  
 
The best fit parameters are shown in Tab.~\ref{tab:fit}.
 We note that spectrum 1 shows spectral parameters of the canonical soft state, according also with the hard X-ray power-law slope (Tab. \ref{tab:log}),
while spectra 9--12 show typical spectral parameters of the hard state (see Zdziarski 2000; Zdziarski \& Gierli\'nski 2004). Spectra 2--8 may be defined as intermediate.
In addition, we call intermediate/soft all spectra where a disc black-body is required by the fitting procedure and intermediate/hard the spectra where 
it is not possible to give constraints  to the inner disc temperature.

\begin{figure}
\includegraphics[height=5.8cm]{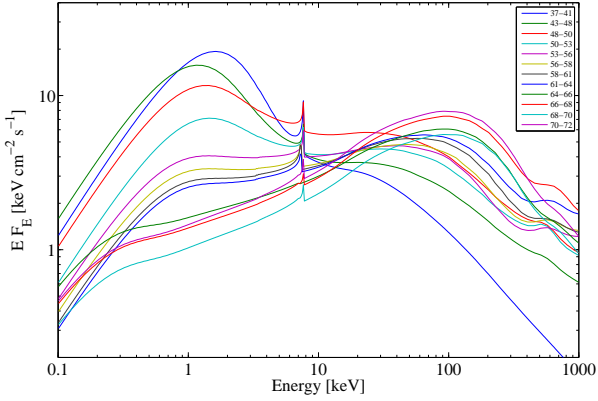}
\caption{Total models obtained by fitting all twelve broad-band spectra with {\sc eqpair} plus {\sc diskline} (see Tab. \ref{tab:fit}) \label{fig:model}}
\end{figure}

In Fig. \ref{fig:corr}, we show the relation between the best-fit model parameters and the hard X-ray  power-law slopes.
Four significantly different energy-spectra of \cyg\ are plotted in Fig. \ref{fig:spectra}, while in Fig. ~\ref{fig:model}  we show the total best-fit models for each of the twelve selected spectra.

In agreement with previous studies (G99) our fits show that the electron-positron 
pairs constitute only a small, if not negligible, fraction of the Comptonising leptons. 
An interesting trend (shown in Fig. \ref{fig:corr}) is that the Thomson optical depth is lower in softest states ($\tau_{\rm p}\simeq0.2$ in SS). 
It increases significantly from SS (spectrum 1) to intermediate states (spectra 2--8) and is approximately constant in the hard states  ($\tau_{\rm p}\simeq1.4$, spectra 9--12).
This change by a factor of 7 could be associated either to a change in electron density  or in the size of the corona (since $\tau\propto n_{e}R$). If this is the size of the corona that changes then this would imply that the corona is larger in the HS than in the SS by a factor of $\sim$7. This would be in agreement with the decrease of the inner radius of the accretion disc  and faster rapid variability in the SS which both suggest  a smaller emitting region in the SS. 
But the  electron density could also drive the change in Thomson depth. This could be the case if the material in the corona gradually condensates into the accretion disc when the source evolves toward softer states. Another possibility would be that the coronal material is evacuated during the ejection event (optically thin radio flare)  at the spectral transition (as suggested by  Rodriguez \& Prat 2010). However we see on Fig.~\ref{fig:corr} that the evolution of $\tau_{\rm p}$ is gradual. It starts  in the intermediate/hard states and continues in the soft state, this appears in contradiction with a scenario in which most of the corona would be suddenly ejected at the transition. If the corona is ejected this has to be a gradual process, that might be associated with the presence of strong disc winds evidenced in the SS of X-ray binaries (see e.g. Ponti et al. 2012).

\begin{figure}
\centering
\includegraphics[height=8.8cm, angle=90]{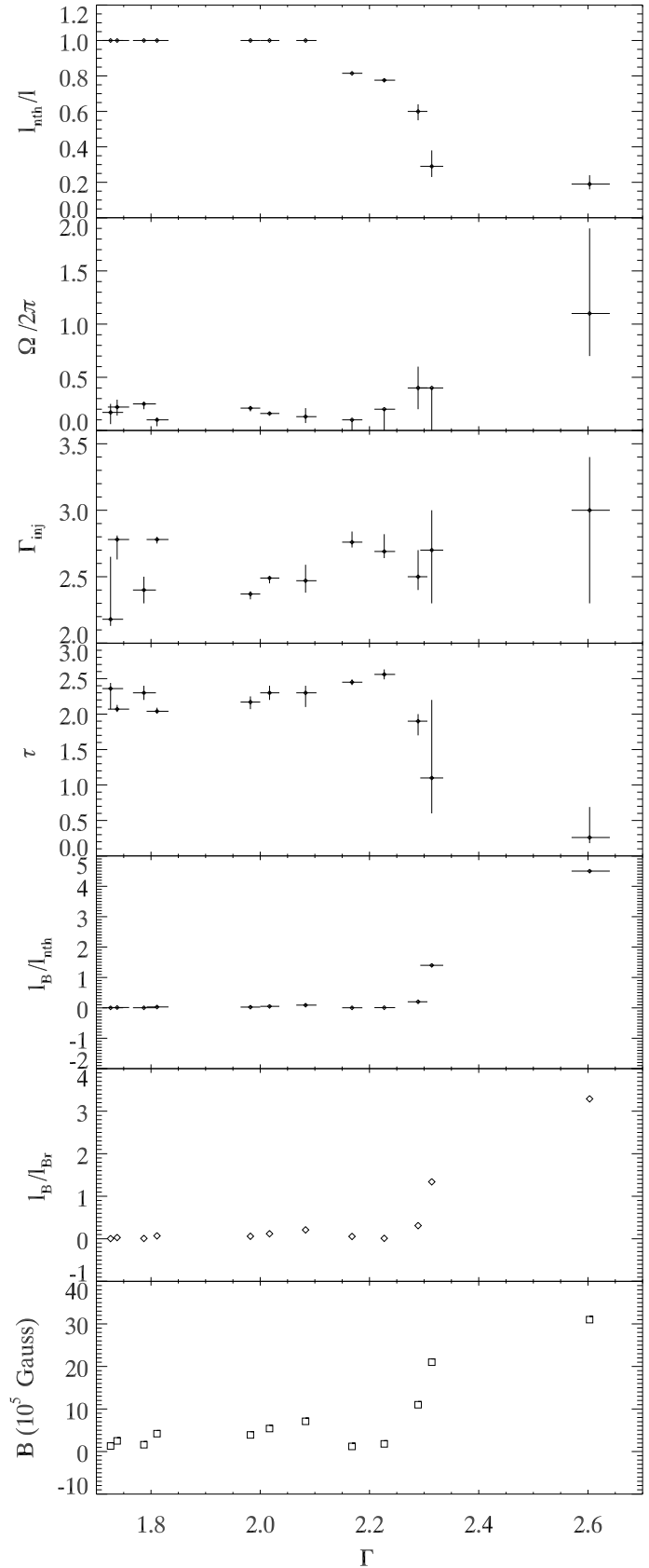}
\vspace{+0.2cm}
\caption{{\sc belm} spectral parameters reported in Tab. \ref{tab:belm} vs $\Gamma$ reported in Tab. \ref{tab:log}.}
\label{fig:corr_belm}
\end{figure}

We find that the inner disc temperature $kT_{\rm in}$ decreases from 0.6 and 0.4 keV  between spectrum 1 and 4 (soft state to intermediate/soft state, see Tab.~\ref{tab:fit}).
This is combined with a decrease by a factor of four of the flux of the measured thermal disc component. For harder spectra, the disc temperature is so low that it  cannot be constrained with our data. 
We therefore imposed an inner disc temperature of  0.3 keV when in intermediate/hard states (spectra 5--8) and 0.1 keV  when in canonical HS (spectra 9--12).  
The temperature of the corona has the opposite behaviour and is larger for harder spectra, except for spectrum 1. 
This change of coronal temperature is driven by a change in the ratio of the electron  heating rate to soft cooling photon flux in the corona ($\propto l_{\rm h}/l_{\rm s}$) which is larger at small 
$\Gamma$. The fact that the temperature appears higher in the softer spectrum 1, is related to the significantly smaller optical depth in soft state. In fact, even if the temperature is higher in spectrum 1, the Compton parameter $y\propto \tau_T T_e$ which is expected to vary approximately like $l_{\rm h}/l_{\rm s}$ is the lowest. Then computing the absolute disc and Comptonized fluxes (respectively $F_{bb}$ and $F_{Compt}$, see Tab.~\ref{tab:fit}) can provide indications on whether this evolution is driven by changes in the flux of soft cooling photons from the disc ($\propto l_{\rm s}$) or changes in the electron heating rate ($\propto l_{\rm h}$).
We find that the thermal disc flux changes by more than one order of magnitude while the Comptonized radiation decreases by almost a factor of two.
The softening we observe in the states 1-4  appears to be caused by a dramatic increase 
in the disc thermal flux  in the corona associated with a modest reduction of the electron heating rate.

Non-thermal acceleration appears to be required to fit most spectra. When trying to fit with a non-thermal compactness fraction fixed to $l_{\rm nth}/l_{\rm h} = 0$,  only for the spectrum 12 we obtained an acceptable $\chi^{2}_{\nu}$ (i.e. 0.98). However, for this last spectrum the statistics is very low  and residuals show that an additional non-thermal component is required at high energy. For the other spectra, significant $l_{\rm nth}/l_{\rm h}$ fractions are required. This indicates 
that a non-thermal electrons population may also be present in the corona (or elsewhere), together with the thermal population even during the hard states.
This is in agreement with previous studies performed on intermediate and hard states of BH binaries, as also reported in Sec. \ref{sec:intro}.
In soft and intermediate/sost states (1--4) though, it was not possible to constrain $l_{\rm nth}/l_{\rm h}$  and this parameter was fixed at 0.99.

From our spectral fits, from the softest (spectrum 1) to the harder states (up to spectrum 9) $\Gamma_{\rm inj}$ parameter changes from 3.5 to 2.0.
We are not able to give constraints on this parameter for spectra 10--12, so that we fixed it to 2. Such a change in the slope of the injected non-thermal electrons suggests a different acceleration mechanism in SS and HS and may support the view that the non-thermal excess in the HS has a different origin (e.g. the jet). We note however that the results of MC02 suggest a different evolution (i.e. $\Gamma_{\rm inj}$ larger in the HS). It is possible that the soft injection we infer in the SS is related to the lack of photon statistic above 100 keV and to the fact that the non-thermal compactnes fraction $l_{\rm nth}/l_{\rm h}$ was fixed when fitting the softest spectra.

 We checked the need for the reflection component freezing $\Omega$/$2\pi$ at 0. 
The results show that the $\chi^2$ are consistently worst (F-test probabilities $<$ 10$^{-4}$), 
confirming that introducing the reflection component improves significantly the fits.
Our fits suggest that the reflection component is significantly higher only in the softest state (spectrum 1) of our data set,
while it is almost constant within the errors in the other states.
This is a known result which can be interpreted as follows. In the HS  the reflecting disc 
intercepts a small  fraction of the coronal radiation since the disc 
is truncated at a large distance from the black hole; 
in the SS the accretion disc extends down to the last stable orbit and is sandwiched by the illuminating corona (Done et al. 2007). 

\subsection{Magnetized, one-zone model with pure non-thermal acceleration}
\label{sec:belm}
\begin{table}
\begin{center}
\caption{Fitting parameters of the {\sc belm} table model used in XSPEC and their properties (minimal and maximal values, number of interpolation values, and grid style). 
Other parameters are set to: $l=10$, $l_{\rm s}=l-l_{\rm nth}$, $l_{\rm th}=0$, $\gamma_{min}=1.3$, $\gamma_{max}=10^3$,  $R=5\times10^7$ cm.}\begin{tabular}{c|cccc} 
\hline
 & min & max & number & grid \\
\hline
$l_{\rm nth}/l$ & 0.1 & 10 & 15 & log \\
$l_B/l_{\rm nth}$ & 0.001 &10 & 15 & log\\
$\tau_{\rm p}$ &0.005& 5 & 15 & log \\
$\Gamma_{\rm inj}$ &1 & 5 & 15 &lin \\
$k T_{\rm in}$ & 20 eV & 2 keV & 10 & log \\
\hline
\end{tabular}
\label{tab:belm_param}
\end{center}
\end{table}

 If the corona is magnetized, it must emit synchrotron emission that can add up to disc photons and be Comptonized. 
In this section we aim at answering the question: how large can be the magnetic field before synchrotron emission makes the modeled spectra incompatible with the data? 
It was shown in MB09 and PV09 that processes such as the synchrotron boiler and Coulomb collisions can thermalize particle distribution efficiently, even when particles are accelerated trough pure non-thermal mechanisms. To minimize the number of parameters, we focus here on such a model with pure non-thermal acceleration. The alternative case of thermal heating will be discussed in the next subsection. 

Using {\sc BELM}, we have computed a table of more than $5\times10^{5}$ spectra and 5 fitting parameters: $l_{\rm nth}/l$, $l_B/l_{\rm nth}$, $\tau_{\rm p}$, $\Gamma_{\rm inj}$, and $k T_{\rm in}$ (see table \ref{tab:belm_param}). This pure SSC model does not include any thermal or Coulomb heating ($l_{\rm th}= l_{c}=0$).
As shown in MB09, the spectrum shape  depends on ratios such as $l_B/l_{\rm nth}$ or $l_{\rm nth}/l$ and not on the total compactness of the source. Therefore the best fit values for these ratios are also insensitive to the exact value of $l$.
The total compactness was thus set to the typical value $l=10$ and $l_{\rm s}$ was computed such as $l_{\rm s}+l_{\rm nth}=l=10$.  

On the contrary, the magnetic field intensity depends on the exact amplitude of the magnetic compactness $l_B$. Once the spectra were fitted and the flux derived from XSPEC, 
the real source compactness $l_{\rm obs}$ was computed using Eq. 1 and assuming a distance of $d=1.86$ kpc. 
The exact magnetic compactness was then scaled linearly to $l_{\rm obs}$ as:
\begin{equation}
l_{B, \rm obs} = \frac{l_{B}}{l_{\rm nth}} \frac{l_{\rm nth}}{l} l_{\rm obs}
\label{eq:lbv}
\end{equation}
where $l_{B}/l_{\rm nth}$ and $l_{\rm nth}/l$ are provided by the fit and insensitive to $l$. The magnetic field $B$ was then estimated from equation (\ref{eq:magnetic}), assuming a size of the
corona as $R=20 R_{g}$.
Alternatively, the observed magnetic field can be compared to the value of magnetic field in equipartition with the radiative energy density. In the limit of low-energy photons ($h\nu<m_{\rm e}c^2$), the corresponding compactness ratio is (Lightman \& Zdziarski 1987):
\begin{equation}
\frac{l_{B}}{ l_{B_{\rm R}}}=\frac{l_{B}}{l_{\rm nth}} \frac{ l_{\rm nth}}{l } \frac{4\pi/3}{1+\tau_{\rm T}/3}
\label{eq:prad}
\end{equation}
This ratio is again insensitive to the choice for $l$ and is also independent of uncertainties on the source size and distance.

\begin{figure}
\centering

\includegraphics[height=6.3cm]{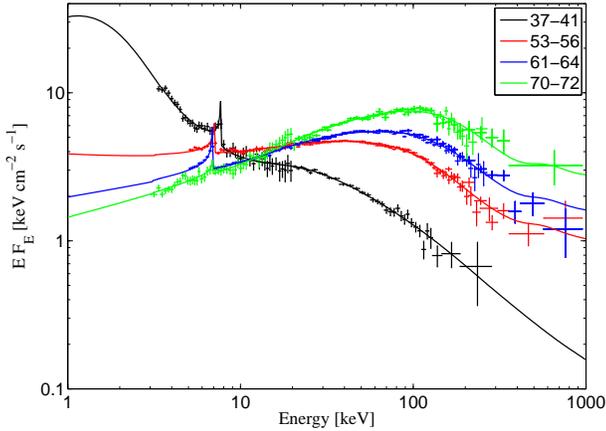}
\caption{Joint JEM-X, IBIS and SPI  energy spectra of \cyg\ during four different spectral states
fitted with the {\sc belm} model with pure non-thermal acceleration
plus {\sc diskline} and {\sc reflect} (see Tab. \ref{tab:belm}).
\label{fig:belm_fit}
}
\end{figure}

\begin{figure}
\centering

\includegraphics[height=6.3cm]{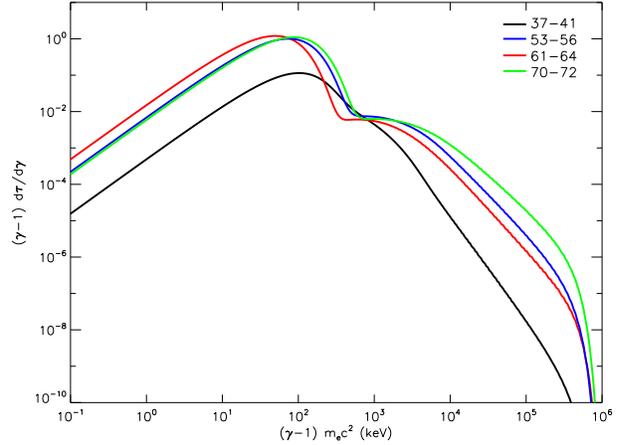}
\caption{Electron energy distributions obtained in the best fit models of Figs.~\ref{fig:belm_mod} and \ref{fig:belm_fit}. The different colours corresponds to those of the spectra shown in Fig.~\ref{fig:belm_fit}.}
\label{fig:belm_leptons}
\end{figure}

\begin{figure*}
\begin{center}
\parbox{16cm}{
\includegraphics[width=0.45\textwidth]{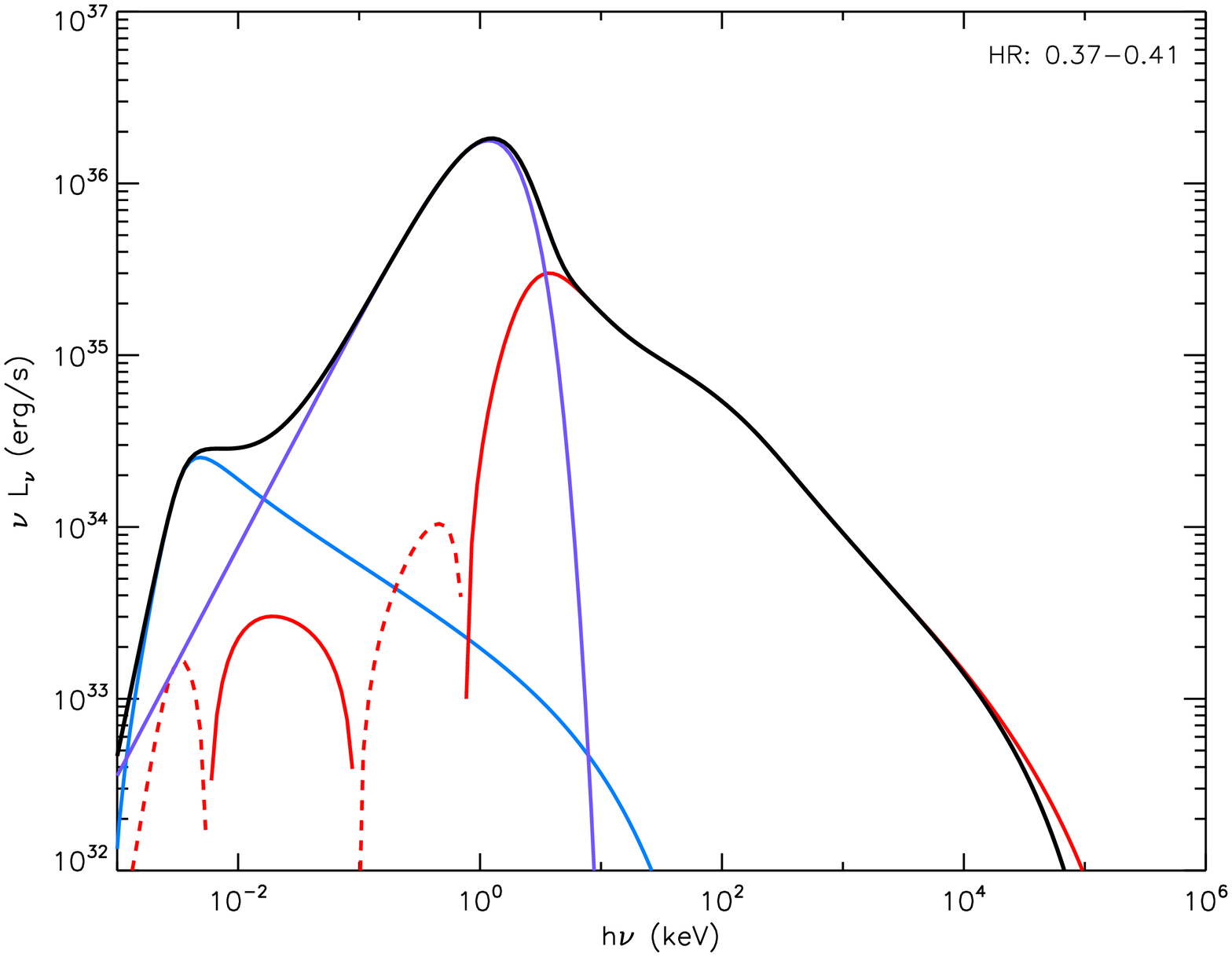}
\includegraphics[width=0.45\textwidth]{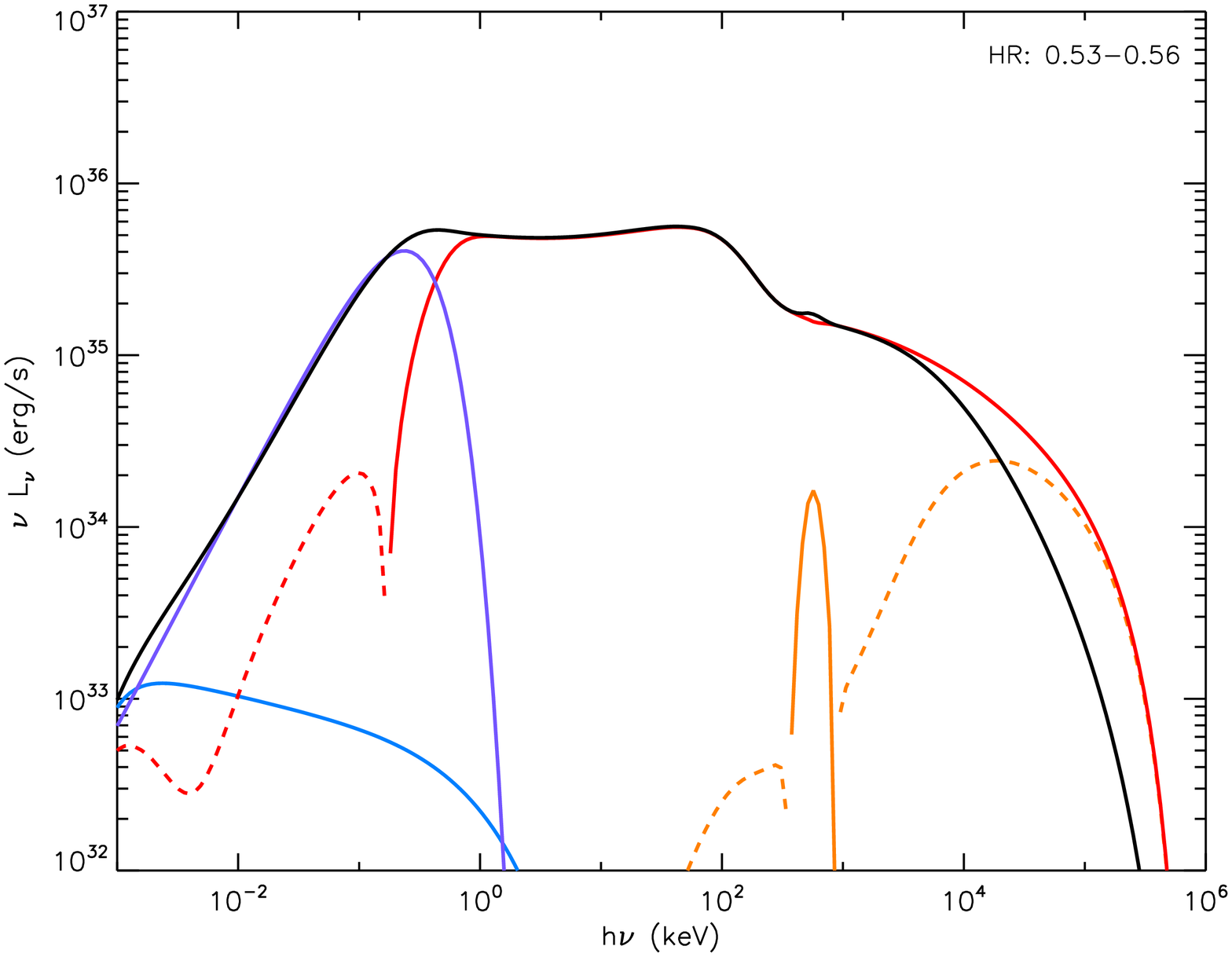}}
\parbox{16cm}{
\includegraphics[width=0.45\textwidth]{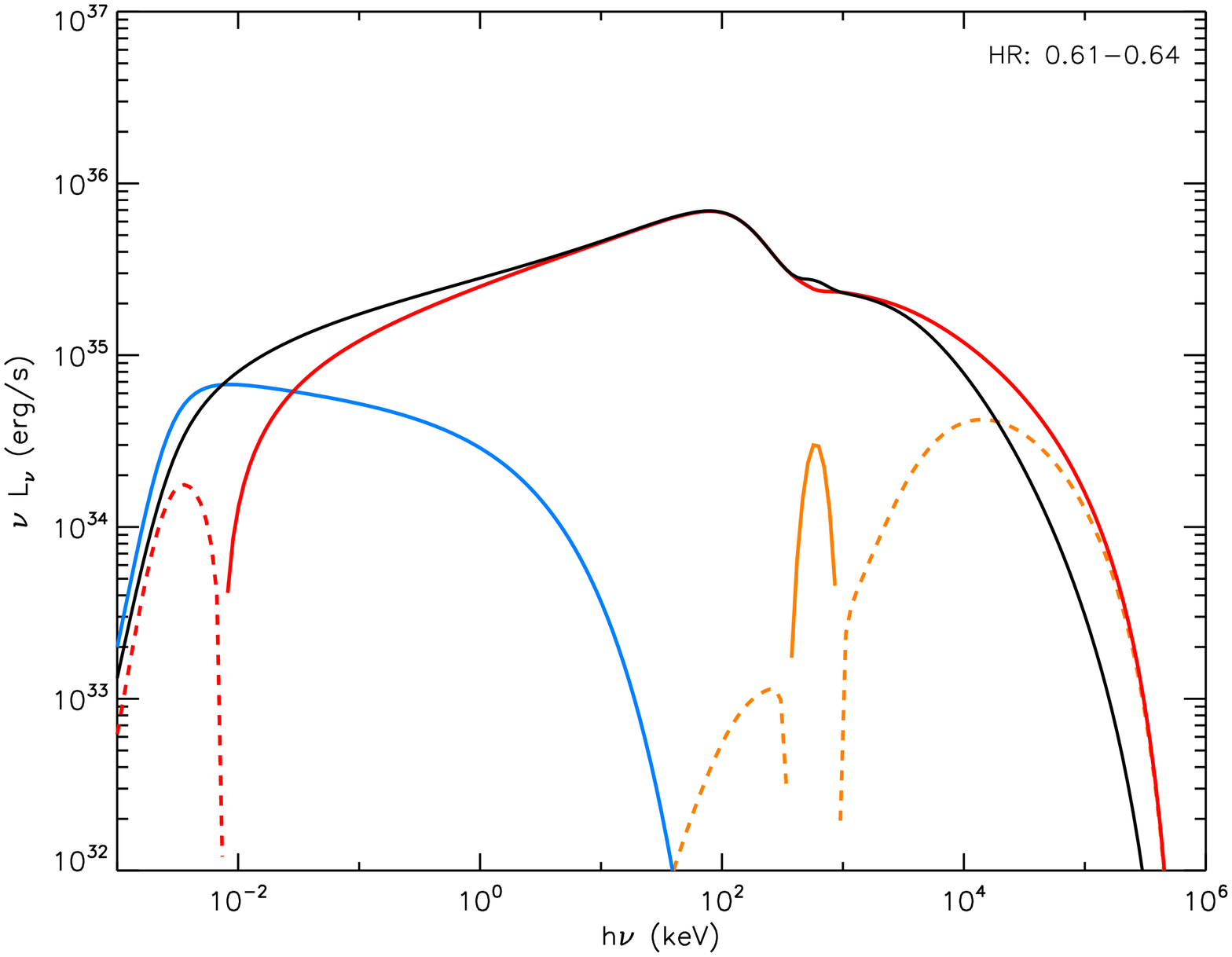}
\includegraphics[width=0.45\textwidth]{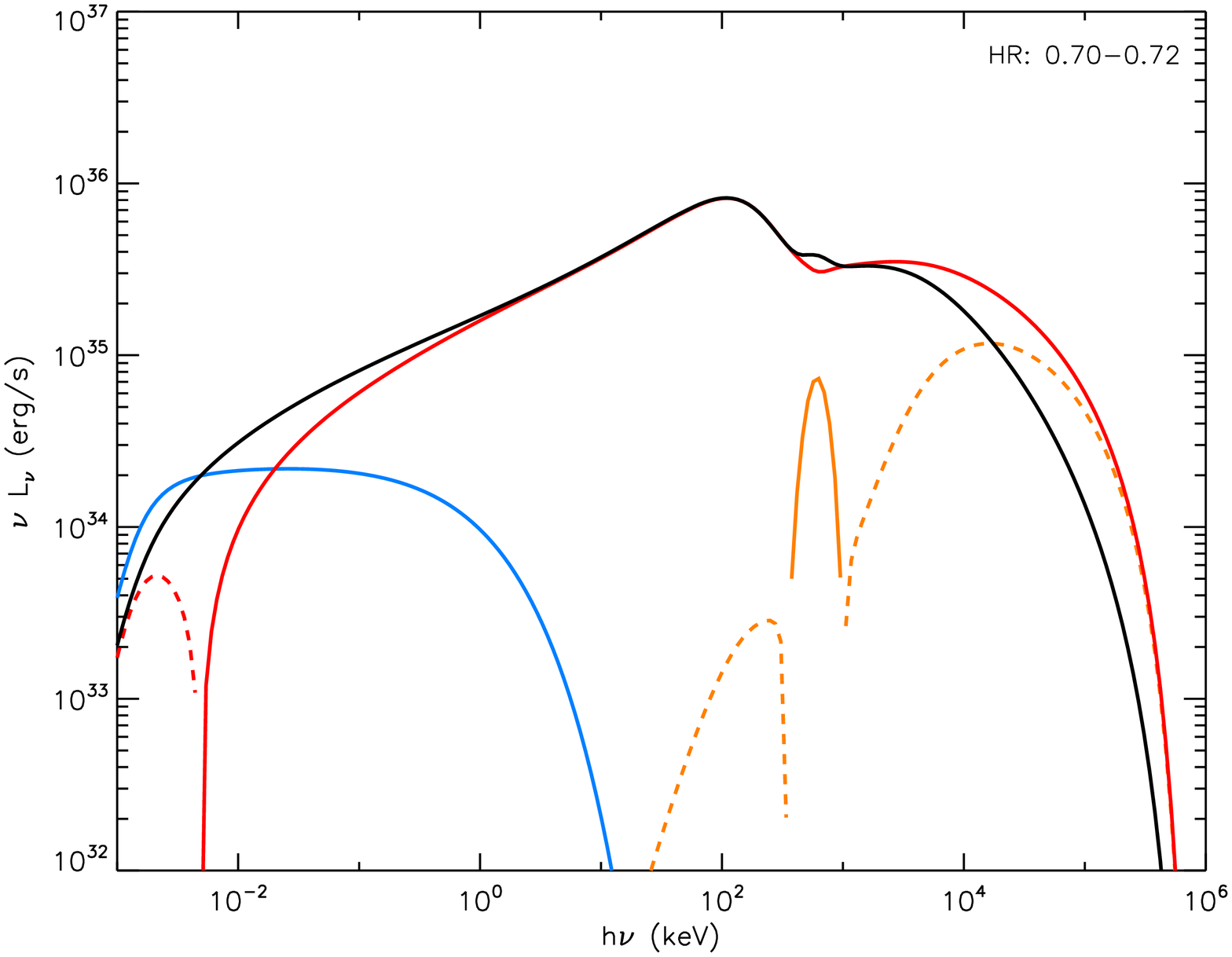}}
\caption{{\sc belm} total models (black) corresponding to  fits of the spectra shown in Fig.~\ref{fig:belm_fit} and the components: Compton (red), pair annihilation/production (orange),
synchrotron (blue), soft black body photons (purple). Solid lines correspond to positive contribution to the spectrum, dashed
lines correspond to negative contribution.}
\label{fig:belm_mod}
\end{center}
\end{figure*}

Tab.~\ref{tab:belm} shows the best-fit parameters. Fig. \ref{fig:belm_fit} presents a selection of four different \integral\ spectra fitted with {\sc belm}, while Figs.~\ref{fig:belm_mod} 
and \ref{fig:belm_leptons} display respectively  the photon spectra and the particle distribution, for the corresponding models.
 Overall this model provide good fit to the data (Tab. \ref{tab:belm}). This confirms that the data are compatible with pure non-thermal acceleration models.

The parameters $\tau_{\rm p}$, $kT_{\rm in}$, $l_{\rm nth}$ (or $l_{\rm h}/l_{\rm s}$) and the reflection amplitude have a similar behaviour as in the fits with the unmagnetized model {\sc eqair}, 
for similar reasons (Fig. \ref{fig:corr_belm}).  The fact that $l_{\rm nth}/l$ is lower  in softer state is equivalent to the trend observed for the ratio $l_{\rm h}/l_{\rm s}$.  We note that when using {\sc belm}
the electron acceleration index $\Gamma_{\rm inj}$ is better constrained by the data in the magnetized model since it  also affects the number of synchrotron emitting electrons which in turns control the temperature of the Maxwellian electrons and the slope of the X-ray spectrum. The trend for a softer injection index in softer state is less clear in the fits with this magnetized model.

From the results of our fits, it is clear that in the soft state, the accretion disc photons represent the main source of soft seed photons (see Fig.~\ref{fig:belm_mod}). In the harder states (spectra 6--12), it is not possible to obtain any constraints and the synchrotron photons may dominate. We have set $l_{\rm s}=0$ when fitting these spectra and obtained good fits. Based on the observation of correlations between X-ray flux and spectral index in several black hole binaries in the HS, Sobolewska et al. (2011) have argued that  synchrotron seeds should dominates at luminosities below about a percent of the Eddington limit, while disc seed photons should dominate at higher luminosities.  It is interesting to note that Cyg X-1 has a stable luminosity located around this transition. It is therefore likely that both the accretion disc and the magnetic field contribute to seed the Comptonization process. 

Nevertheless, since it is not possible to constrain their respective contribution from our data, assuming the extreme case of pure synchrotron seed photons allows us to set an upper limit on the magnetic compactness. Indeed the total flux of seed photons (synchrotron+disc) determines the slope of the X-ray spectrum (via energy balance of the thermal electrons).  Since the flux of synchrotron photons increases with $l_{B}$, and the spectral slope is well constrained by the data, assuming that there are no disc photons is going to maximize $l_{B}$. In the softer states the spectra are consistent with $l_B=0$  (since they can be fit with  an unmagnetized model) but again we are able to obtain an upper limit on $l_B$.

We note that the jump in $l_{B}/l_{\rm nth}$ and $B$ between spectra 5 and 6 is due to the fact that we had to  turn off the disc seeds photon when fitting spectra 6--12. This makes the upper limit on $l_B$ less restrictive. The presence of this jump suggest that disc photons continue to play a role even in the harder spectra. 

From Tab.~\ref{tab:belm}, we see that for a corona of size $R=20R_g$, these upper limits range from $\simeq10^{5} $ G in the hardest HS to $\simeq 3\times 10^{6}$ G 
in the soft state (see also Fig. \ref{fig:corr_belm}). 
This implies strongly subequipartition magnetic fields ($l_B/l_{B_R}<1$) in the spectra intermediate and hard. As discussed in MB09 and Droulans et al. (2010) , this has important consequences for dynamical models of the corona. 
Namely, this rules out any model based on a X-ray corona powered through magnetic dissipation in the hard state and this may also be a problem to drive the powerful jet that is observed.

\subsection{Magnetized, two-zone model with pure thermal heating}
In the previous one-zone model, we obtained upper limits on the magnetic field intensity  assuming Synchrotron Self-Compton emission (SSC) and pure non-thermal acceleration. This was motivated by the overall non-thermal nature of the softer spectra and by the high energy emission above the cutoff in the harder states (HR$>$0.64),  none of which is reproduced with purely thermal lepton populations.
 In the hard state, these constraints were mostly governed by the non-thermal, highest energy data points.
However, it has been suggested that the high energy emission in the hard states could originate from a different region of the accreting system, i.e. the jet \cite{laurent11,zdz12} or another component of the corona (Malzac 2012). 
Then, the hard spectra  below the cutoff could also be reproduced by SSC models with pure thermal heating. In this case the exact heating mechanism is of little importance for spectral modeling, because whatever the heating mechanism the resulting electron distribution is very close to a pure Maxwellian.  In {\sc belm} this electron thermal heating mechanism is assumed to be Coulomb collisions with a distribution of hot thermal protons (MB09).  
Then, the constraints on the magnetic field are expected to be different.

To investigate this issue, we computed a new table model with pure Coulomb heating ($l_{\rm s}=0$, $l_{\rm nth}=0$, $l=l_c=l_{\rm th}$). As before, the spectral shape does not depend on the absolute compactness of the system, and we set $l=l_{\rm c}=1$ to prevent degeneracy issues. Then we determined the proton temperature ($kT_{\rm p}$ in Tab. \ref{tab:coulomb}) corresponding to this compactness and to the exact steady state electron distribution.
 The two free parameters of the table model are: $l_B$ (20 values between 0.1 and $10^4$), 
and $\tau$ (20 values between 0.5 and 5). As for previous models, we also added a reflection component.
We fitted the four harder spectra (HR$>$0.64; see Tab. \ref{tab:coulomb}).

As expected, the broadband spectrum can not be fitted with this pure thermal model, since it does not reproduce the highest energy data, which either requires a significant level of non-thermal acceleration or must originate in a different region. When excluding data above 200 keV, we obtained good fits with $\chi^2_\nu < 0.95$. The results are shown in Tab. \ref{tab:coulomb}. The best fit Thomson optical depth and the reflection amplitude are comparable to those of the non-thermal model.

Noticeably, the ratio $l_B/l_{c}$ is larger by about 4 orders of magnitude in the thermal model than in the previous non-thermal model. Indeed, the spectrum below the cutoff is essentially governed by the temperature of the thermal electron distribution. The cooling of thermal electrons is dominated by inverse Compton scattering on soft synchrotron photons. Non-thermal models produce stronger high energy tails in the electron distribution and thus tend to provide more synchrotron seed photons than thermal models for comparable $l_B$. Hence the magnetic field has to be weaker in non-thermal models in order to provide the same electron temperature and reproduce the data.

Thus, by assuming that the highest energy emission originates from a different region, the constraints on the ratio $l_B/l_{c}$ estimated in this thermal model correspond to the most conservative upper limits. Following the procedure described in the previous subsection, we obtain that the magnetic field in the hard states is stronger by a factor of 100 compared to that estimated in the SSC model (Tab. \ref{tab:coulomb}).  The magnetic field is now allowed to be strongly super equipartition in the hard state with typical upper limits  of the order of $10^{7}$ G.

In the case of a pure Maxwellian plasma, the dependence of the Coulomb compactness ($l_c$) on electron and proton temperatures can be re-written as:

\begin{equation}
\frac{T_{\rm p}}{T_{\rm e} }\simeq 1+7\frac{gl_{\rm c (obs)}}{\tau^{2}},
\label{eq:temp}
\end{equation}
where $g(\theta_{\rm e})=\sqrt{7\theta_{\rm e}}$ and $\theta_{\rm e}=\frac{kT_{\rm e}}{m_{\rm e}c^2}$ is the electron temperature in units of the electron rest mass (see MB09 for details).   $l_{\rm c(obs)}=l_{\rm obs}$ is obtained from equation (\ref{eq:comp}) and the bolometric fluxes by the fits (Tab. \ref{tab:coulomb}). We see that temperature of the ions is at most of a few MeV which is incompatible with standard two temperature accretion flow models (such as ADAF models). As discussed in MB09, this relatively low proton temperature is related to the large (i.e. $\ga$1) Thomson optical depth required to fit the data. As discussed in Malzac (2012), standard hot disc solution do not produce such large Thomson depth (and low ion temperature). The magnetically  dominated accretion flow model of Oda et al. (2010, 2011) may however produce such parameters. The strong magnetic field assumed in this model would be consistent with our upper limits only in the case of a pure thermal electron plasma with the MeV tail produced in the jet or in an other region of the corona.

\section{Conclusion}

We have used data from  6 years of observation of Cygnus X-1 with \integral, to produce 12 high quality,  stacked broad-band spectra spanning the whole range of spectral shapes observed in this source from the SS to HS. We have then fitted these spectra with hybrid thermal/non-thermal Comptonization models. We have used both a magnetized {\sc belm} and a unmagnetized {\sc eqpair} models.
For both models we found that the spectral changes at high energy are driven essentially by the strength of the soft seed cooling photons in the corona as the disc temperature (and inner radius of emission) changes. 

During the spectral transition from HS to SS  the power of the corona decreases by less than a factor of 2 while the disc emission increases at least by a factor of $\sim 50$.  This appears to be associated with a gradual change of the Thomson depth of the corona which decreases from $\simeq1.5$ in HS to $\simeq 0.2$ in SS, that may be caused either by a condensation of the corona into the disc or evaporation into an outflow, or compactification of the corona in the SS. 

For the first time we have performed a statistical fit of the \cyg\ data with the {\sc belm} model. This allowed us to test the effects of the magnetic field on the coronal emission. 
As explained in the Sec. \ref{sec:belm},  the estimate of B depends only on our choice for R and not on the assumed fixed value of $l$ used in the fitting procedure. Within the one-zone model the uncertainty on R is by far the main source of error on our estimate of B, and the other parameters have none or very little effects.

We found that in the softer states, the emission is dominated by Comptonization of the disc photons and the upper limit on the magnetic field is at most of the order of 3 $\times 10^{6} \sqrt{20 R_g/R}$ G. 
Such a magnetic field is energetically consistent with a magnetic dissipation process, such as reconnection, powering the corona. 

In the hard states the data are consistent with a pure  Comptonized synchrotron emission model, although Comptonization of disc photons cannot be excluded.  We consider that a mixture of both is the most likely possibility. However, fitting in the limiting case of SSC in the HS provides upper limits on the magnetic field intensity.  

If the non-thermal excess observed above a few hundred keV is produced in the same region as the bulk of the thermal Comptonization, the magnetic field is at most of a few $10^{5} \sqrt{20 R_g/R}$ G. This is strongly subequipartion with radiation which is challenging for all dynamical accretion flow models. In this case, we also confirm the qualitative results of MB09 and PV09  that the emission in all spectral states can be modeled with pure non-thermal electron acceleration. Indeed, provided that the magnetic field is not too strong in the HS, the electrons thermalise at the observed temperature under the effects of Coulomb collisions and synchrotron self-absorption.  
Moreover, Wardzi\'nski et al. (2002)  found analytically that if the high-energy tail observed in HS of BHBs is from the hybrid distribution, the magnetic field
has to be strongly subequipartition, similar to our finding.

If, on the other hand the non-thermal excess is produced in a different location, the constraints on the magnetic field in the hard state are somewhat relaxed and we obtain an upper limit  of about $2\times 10^{7} \sqrt{20 R_g/R}$ G in the hard state. 
Such a large magnetic field would be consistent with the magnetically dominated hot accretion flow model of Oda et al. (2010) which also appear to produce the rather large Thomson depth ($\tau_{\rm p}\simeq 1.5$) and low ion temperatures $T_{\rm p}\sim10^{10}$ K, that we infer from the data in the hard state.

\section*{Acknowledgments}
MDS and GDC acknowledge financial support  from the
agreement ASI-INAF I/009/10/0 and from PRIN-INAF
2009 (PI: L. Sidoli). MDS and JM acknowledge financial support from Universit\'e Paul Sabatier.
JM \& RB acknowledge support from PNHE in France. MDS thanks IRAP/CNRS (France) for hospitality. JM thanks the Institute of Astronomy of Cambridge (UK) for hospitality. 
MDS and JM thank A. Zdziarski for useful scientific discussion and suggestions.
MDS thanks J. Wilms  and V. Grinberg for useful comments.


\begin{landscape}
\begin{table}
\begin{center}
\caption{The fit parameters of the joint JEM-X, IBIS and SPI spectra in twelve different spectral states.
Fits have been performed with {\sc eqpair} combined with {\sc diskline}. Values in parenthesis denote parameters fixed during the fits.
One of the interesting fixed parameters is $l_{\rm s}$ ($l_{\rm bb}$ in  {\sc eqpair}) which is
fixed at 10 for states 1-4 and at 1 for states 5-12. The inner disc temperature is fixed at 0.3 keV in the intermediate states, while it is 0.1 keV in the harder ones.
See text for the parameters description.}\label{tab:fit}
\renewcommand{\arraystretch}{1.3}
\begin{tabular}{cccccccccccccc} 
\hline
\hline
State & HR   &  $l_{\rm h}/l_{\rm s}$ & $l_{\rm nth}/l_{\rm h}$&$\tau_{\rm p}$& $kT_{\rm in}$ &$\Omega$/${2\pi}$& $\Gamma_{\rm inj}$&$\tau_{\rm tot}$ &$kT_{\rm e}$&$\chi^2_{\nu}$(dof) & 
\multicolumn{3}{c}{\rm Flux $\times 10^{-8}$} \\	
      &        &                   &            &        &     [eV]             &         &             &  &   [keV]      &   & \multicolumn{3}{c}{[\ergcms]} \\
      &        &                   &            &        &                      &         &             &  &              &   & $Bol$ & $bb$& $Compt$  \\
\hline

1 & 0.37-0.41&$0.30^{+0.05}_{-0.04}$ & (0.99) & $0.20^{+0.18}_{-0.08}$ & $627^{+31}_{-108}$ & $0.9^{+0.3}_{-0.1}$ & $3.44^{+0.07}_{-0.08}$ &0.20 & 48.0 & 0.82(89)& 8.1 & 5.2 & 2.4 \\
2& 0.43-0.48 &$0.7 \pm 0.1$ & (0.99)  & $0.8^{+0.2}_{-0.4}  $  & $416_{-39}^{+144}$ & $0.4^{+0.2}_{-0.1} $   & $2.93^{+0.04}_{-0.06}$  & 0.86 & 25.0 & 0.97(127)& 7.8 & 3.4 & 4.2 \\
3&0.48-0.50&$1.35^{+0.08}_{-0.20}$ & $ (0.99)$ & $1.21^{+0.15}_{-0.08}$ & $426^{+93}_{-75}$& $0.41^{+0.08}_{-0.16}$& $2.92^{+0.05}_{-0.02}$ &1.24 & 25.7 & 0.99(129)&7.8  & 2.1 & 5.2 \\
4&0.50-0.53 &$1.63^{+0.04}_{-0.02}$ & (0.99)&$1.44^{+0.01}_{-0.03}$& $430^{+24}_{-20} $ &  $0.37^{+0.03}_{-0.02} $& $2.80^{+0.03}_{-0.02}$ &1.48 & 23.7 & 1.04(125)& 5.4 & 1.2   & 3.8  \\
5&0.53-0.56&$3.88^{+0.09}_{-0.06}   $ & $0.75^{+0.06}_{-0.04}  $ &  $ 1.63 \pm 0.02  $   & $(300)$  & $0.17 \pm 0.03$  &$ 2.37 ^{+0.08}_{-0.03} $  & 1.7 & 37.6 & 0.98(120)& 4.6  & 0.6  & 3.8  \\
6&0.56-0.58&$4.4^ {+0.2}_{-0.1} $ & $0.83 \pm 0.08 $ &  $ 1.61^{+0.20}_{-0.03} $   & $(300)$  & $0.28^{+0.02}_{-0.04} $  &$ 2.4 \pm 0.1 $  & 1.62 & 41.0 & 0.98(116)& 4.3 & 0.5  & 3.5 \\
7&0.58-0.61&$5.19^{+0.06}_{-0.15}  $ & $0.87^{+0.05}_{-0.02}$ &  $1.51^{+0.03}_{-0.02} $   & $(300)$  & $0.29^{+0.03}_{-0.01}$  &$ 2.50^{+0.03}_{-0.17}$  & 1.52 & 47.2 & 0.99(118)& 4.3  & 0.4 & 3.6  \\
8&0.61-0.64&$6.1^{+0.1}_{-0.2}$ &$0.86^{+0.06}_{-0.05} $  &  $ 1.50^{+0.07}_{-0.03} $ & $ (300) $ & $0.32 \pm 0.03$  & $ 2.33^{+0.03}_{-0.06} $   & 1.53 & 48.4 &  1.09(119)&4.4 & 0.4 & 3.6 \\
9&0.64-0.66&$11.3^{+0.3}_{-0.2}$ &$0.55^{+0.01}_{-0.07} $  &  $ 1.30^{+0.02}_{-0.06} $ & $ (100) $ & $0.23^{+0.04}_{-0.02}$  & $  2.5^{+0.4}_{-0.3}  $   & 1.34 & 67.6&  0.91(125)&  4.2 & 0.3 & 3.8 \\
10&0.66-0.68&$16.2^{+0.2}_{-0.4}$ &$0.51\pm  0.03 $  &  $ 1.38^{+0.06}_{-0.04} $ & $ (100) $ & $0.36^{+0.11}_{-0.02}$  & $(2.0) $   & 1.53 & 64.6&  0.96(66)&4.7 & 0.1& 4.0 \\
11&0.68-0.70&$16.3^{+0.8}_{-0.3} $ &$ 0.33^{+0.02}_{-0.03}  $  &  $ 1.30^{+0.06}_{-0.02} $ & $ (100) $ & $0.31^{+0.03}_{-0.02}$  & $(2.0) $   & 1.41& 76.0 &  1.04(130)&3.5 & 0.1 & 3.1 \\
12&0.70-0.72&$15.8^{+0.7}_{-1.2}$ &$ 0.3^{+0.2}_{-0.1} $  &  $ 1.3 \pm 0.1 $ & $ (100) $ & $0.33^{+0.06}_{-0.04} $  & $(2.0) $   & 1.39 & 77.2 &  0.95(132)&4.9 & 0.1 & 4.3 \\

\hline
\end{tabular}
\vspace*{3 cm} 
\end{center}
\end{table}
\end{landscape}


\begin{landscape}
\begin{table}
  \begin{center}
    \caption{Same spectra than in Tab. \ref{tab:fit} fitted with the {\sc BELM} model with pure non-thermal acceleration plus {\sc diskline} and {\sc reflect}.}\label{tab:belm}

\renewcommand{\arraystretch}{1.3}
\begin{tabular}{ccccccccccc} 
\hline
\hline
HR   &  $l_{\rm nth}/l$ & $l_{B}/l_{\rm nth}$&$\tau_{\rm p} $& $\Gamma_{\rm inj}$ &$kT_{\rm in}$ &$\Omega$/${2\pi}$&$\chi^2_{\nu}$(dof) & $\rm F_{Bol} \times 10^{-8}$  & $\rm B$ & $l_{B}/l_{Br}$\\	
        &                          &                                    &                    &                 &  [keV]        &                                &                                                & [\ergcms]      & [Gauss]     &                       \\
\hline
0.37-0.41&$0.19^{+0.05}_{-0.03}$ &  $< 4.5 $ & $0.26^{+0.43}_{-0.08}$ & $ 3.0^{+0.4}_{-0.7}$ & $ 0.5\pm 0.2 $& $ 1.1^{+0.8}_{-0.4}$   &  0.85(88) & 7.2 & $3.1 \times 10^{6}$ & 3.29 \\
0.43-0.48&$0.29^{+0.09}_{-0.06}$ &  $< 1.4 $ & $1.1^{+1.1}_{-0.5}$ & $ 2.7^{+0.3}_{-0.4}$ & $ 0.21^{+0.06}_{-0.04}$& $ <0.4$   &  0.96(126)& 6.7   & $2.1 \times 10^{6}$ & 1.34 \\
0.48-0.50&$0.60^{+0.04}_{-0.05}$ &  $<0.2 $ & $1.9^{+0.1}_{-0.2}$ & $ 2.5^{+0.2}_{-0.1}$ & $ 0.21^{+0.06}_{-0.04} $& $ 0.4\pm0.2$   &  1.05(128) & 6.3 & $ 1.1 \times 10^{6}$ & 0.31\\
0.50-0.53&$0.776^{+0.008}_{-0.005}$ &  $<0.007 $ & $2.56 \pm 0.07$ & $ 2.69^{+0.13}_{-0.05}$ & $ (0.2) $& $ < 0.2 $   &  1.0(112)& 3.8 & $1.8 \times 10^{5}$ & 0.012\\
0.53-0.56&$0.815^{+0.004}_{-0.002}$ &  $<0.003$ & $2.45\pm 0.04$ & $ 2.76^{+0.08}_{-0.04}$ & $ (0.1) $& $ < 0.1 $   &  1.1(120)& 3.8  & $1.2 \times 10^{5}$ & 0.056\\
0.56-0.58&$  (1) $ &  $0.09^{+0.05}_{-0.02}$ & $2.3^{+0.1}_{-0.2}$ & $ 2.47^{+0.12}_{-0.09}$ & $ - $& $ 0.13^{+0.08}_{-0.06}$   &  0.94(117)& 3.5 & $7.1 \times 10^{5}$ & 0.21  \\
0.58-0.61&$  (1) $ &  $0.050^{+0.002}_{-0.006}$ & $2.3 \pm 0.1$ & $ 2.49^{+0.01}_{-0.04}$ & $ - $& $ 0.16 \pm 0.01$   &  0.99(119)& 3.7  &  $5.4 \times 10^{5}$ & 0.12\\
0.61-0.64&$  (1) $ &  $0.025 \pm 0.001$ & $2.17^{+0.08}_{-0.10}$ & $ 2.37^{+0.02}_{-0.04}$ & $ - $& $ 0.21^{+0.02}_{-0.03} $   &  0.93(117)& 3.8  &  $3.9 \times 10^{5}$ & 0.06\\
0.64-0.66&$  (1) $ &  $0.03^{+0.02}_{-0.01}$ & $2.04^{+0.05}_{-0.03}$ & $ 2.78^{+0.02}_{-0.03}$ & $ -  $& $ 0.10^{+0.02}_{-0.06} $   &  0.85(126)& 3.6  &  $4.2 \times 10^{5}$ & 0.07\\
0.66-0.68&$  (1) $ &  $0.004^{+0.002}_{-0.001}$ & $2.3 \pm 0.1$ & $ 2.4 \pm 0.1$ & $ -  $& $ 0.25^{+0.02}_{-0.05} $   &  1.07(67)& 4.2  &  $1.6 \times 10^{5}$ & 0.009\\
0.68-0.70&$  (1) $ &  $0.012 \pm 0.002$ & $2.07^{+0.06}_{-0.04}$ & $ 2.78^{+0.03}_{-0.15}$ & $ - $& $ 0.22^{+0.07}_{-0.08} $   &  0.96(118)& 3.2 &  $2.5 \times 10^{5}$ & 0.03\\
0.70-0.72&$  (1) $ &  $0.0025^{+0.017}_{-0.001}$ & $2.36^{+0.08}_{-0.29}$ & $ 2.18^{+0.47}_{-0.05}$ & $ -  $& $ 0.17^{+0.08}_{-0.11} $   &  1.03(132)& 4.6  &  $1.3 \times 10^{5}$ & 0.006\\
\hline
\end{tabular}
\vspace*{3 cm} 

 \end{center}
 \end{table}
\end{landscape}


\begin{table*}
  \begin{center}
    \caption{Spectra below 200 keV in HS fitted with a pure thermal heating model. $kT_{\rm p}$  and $kT_{\rm p(obs)}$ are the temperatures of the ions for $l_{\rm c} = 1$
    and $l_{\rm c}$ observed ($l_{\rm c} = l$), respectively.}\label{tab:coulomb}

\renewcommand{\arraystretch}{1.3}
\begin{tabular}{ccccccccccc} 
\hline
\hline
HR   &  $ l_{B}/l_{c} $  &   $\tau_{\rm p} $   &   $\Omega$/${2\pi}$  &   $\chi^2_{\nu}$(dof) & $\rm F_{Bol} \times 10^{-8}$ & $B$        & $l_{B}/l_{Br}$& $kT_{\rm e}$&$kT_{\rm p}$ &$kT_{\rm p(obs)}$\\
        &                          &                         &                                       &                                     &               [\ergcms]                         & [Gauss]     &                                    &   [keV]   & [keV] &     [MeV]         \\
\hline
0.64-0.66 & $68^{+12}_{-3}$ & $ 1.83^{+0.2}_{-0.02}  $& $ 0.12^{+0.05}_{-0.04} $   &  0.94(107)  &  3.5& $1.9 \times 10^{7}$ & 176.3 & 79.0 &245& 1.6\\
0.66-0.68&$ 44^{+4}_{-8}$ & $ 2.0 \pm 0.1 $& $ 0.27^{+0.01}_{-0.04} $   &  0.94(58)& 3.9   &  $1.7 \times 10^{7}$ & 110.3  & 70.2   & 219 & 1.5 \\
0.68-0.70&$ 48^{+12}_{-4}$ & $ 2.0^{+0.2}_{-0.06}  $& $ 0.19^{+0.05}_{-0.06} $   &  0.90(112) &3.0 &  $1.5 \times 10^{7}$ & 119.8  &  78.2  & 215& 1.1\\
0.70-0.72&$  64^{+17}_{-21}$ & $ 2.1^{+0.3}_{-0.2}  $& $ 0.14^{+0.12}_{-0.06} $   &  0.95(113) & 4.1 &  $2.0 \times 10^{7}$ & 157.1&    72.3& 183  & 1.3\\

\hline
\end{tabular}
\vspace*{3 cm} 
 \end{center}
 \end{table*}

\end{document}